\documentclass[aps,prl,twocolumn,floats, groupedaddress]{revtex4-2} 
\usepackage{amsmath}
\usepackage{graphicx}
\usepackage{dcolumn}
\usepackage{bm}
\usepackage{amssymb}
\usepackage{latexsym}
\usepackage{color}
\usepackage{subfigure}
\usepackage{ulem}
\usepackage{hyperref}

\usepackage{cancel}

\begin{document}

\title{Is the Aharonov-Casher phase geometrical or dynamical?}

\author{Igor Kuzmenko$^{1,2}$, Y. B. Band$^{1,2,3}$, Yshai Avishai$^{1,4}$}

\affiliation{
  $^1$Department of Physics,
  Ben-Gurion University of the Negev,
  Beer-Sheva 84105, Israel
  \\
  $^2$Department of Chemistry,
  Ben-Gurion University of the Negev,
  Beer-Sheva 84105, Israel
  \\
  $^3$The Ilse Katz Center for Nano-Science,
  Ben-Gurion University of the Negev,
  Beer-Sheva 84105, Israel
  \\
  $^4$Yukawa Institute for Theoretical Physics, Kyoto, Japan
  }

\begin{abstract}
We consider two two-dimensional (2D) electronic systems in the presence of a perpendicular homogeneous electric field that generates a Rashba spin-orbit interaction (RSOI): a system of non-interacting electrons in a 2D conductor, modeled using  the 2D Schr\"odinger equation (SE), and a single-layer graphene system, modeled using a 2D Dirac equation (DE) for massless fermions.  In both cases the RSOI is expressed via an $SU(2)$ Rashba vector potential ${\bf A}_{R}$.  We demonstrate that ${\bf A}_{R}$ cannot be eliminated from either the 2D SE or the 2D DE via a gauge transformation.  Nevertheless, for a plane wave solution, an $SU(2)$ matrix exists that eliminates ${\bf A}_{R}$ from the resulting 1D SE.  This unitary matrix is an Aharonov-Casher (AC) phase factor, and facilitates the calculation of the AC phase in the Schr\"odinger scheme.  The plane wave solution for the DE contains two components of ${\bf A}_{R}$: $A_{R, k}$ in the direction of the wave vector ${\bf k}$, and $A_{R, n}$ normal to ${\bf k}$.  The latter generates an effective electron mass that cannot be eliminated from the DE.  The former generates an AC phase that can be eliminated by a time-dependent unitary transformation.  Thus, the Dirac AC phase is time-dependent, i.e., it is a dynamical phase.  This is in contradistinction to the Schr\"odinger AC phase which is geometrical.
\end{abstract}

\maketitle

{\it Introduction}---The Rashba effect \cite{Rashba_59}, a momentum-dependent spin-splitting of energy bands, is observed in a wide range of condensed matter materials. These include non-centrosymmetric wurtzite semiconductors such as CdS and CdSe \cite{Casella_60}, semiconductor heterostructures such as quantum wells (InGaAs/InAlAs) \cite{Nitta-97, Bergsten_06}, surface Shockley states of heavy metals by spin-,angular-resolved photoemission spectroscopy \cite{LaShell_96, Hoesch_04}, bulk BiTeI which shows a giant Rashba spin-orbit splitting \cite{Ishizaka_11}, ferroelectric GeTe \cite{DiSante_12}, hexagonal magnesium-based chalcogenide monolayers (MgX, X = S, Se, Te) \cite{Mohanta-2-21}, quasi-two-dimensional electron gas at the LaAlO$_3$/SrTiO$_3$ interface \cite{Caviglia-2010}, oxide heterostructures of LaAlO$_3$/SrTiO$_3$/LaAlO$_3$ \cite{Lin-2019}, a topological insulator Bi$_2$Se$_3$ ultra-thin film grown on a transition metal dichalcogenides MoTe$_2$ substrate \cite{Wang-2017}, a topological insulator Bi$_2$Se$_3$ \cite{Zhu-2011}, and in atomic systems, see Ref.~\cite{Sangster_1993} which presents measurements of the Aharonov-Casher (AC) phase for two coherent molecular beams of thallium fluoride which have opposite magnetic moments and are {\it not} spatially separated and pass through the same electric field.  Reference~\cite{Nagasawa_12} observed AC oscillations in  InAlAs/InGaAs two-dimensional electron gas rings that depend on the Rashba spin-orbit interaction (RSOI) strength, and Refs.~\cite{Manchon-2015, Mohanta_26} list materials where the RSOI and also where Dresselhaus SO (conduction band energy band splitting in non-centrosymmetric crystals, such as III-V semiconductors, e.g., GaAs, InAs, with zinc-blende structures due to bulk inversion asymmetry) were observed, and Weyl Hamiltonians have been used to model specific condensed matter materials \cite{Armitage_18}.

The occurrence of RSOI gives rise to the AC effect \cite{AC_84} as described below. In particular, we address the question of whether the $SU(2)$ AC vector potential can be eliminated via a suitable gauge transformation.

\begin{figure}
\includegraphics[width=0.9\linewidth,angle=0] {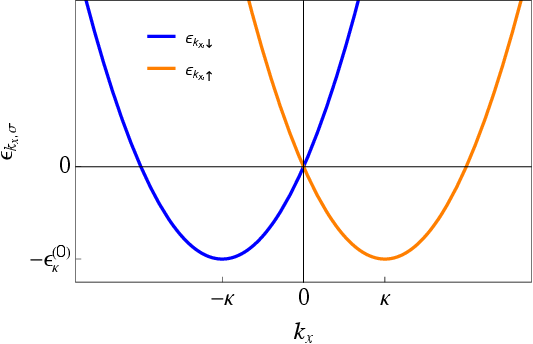}
\caption{\footnotesize
Eigenenergy $\epsilon_{k_x,\sigma}$ of the Schr\"odinger Hamiltonian in Eq.~(\ref{eq:H_S-2D-square}) with RSOI, defined in Eq.~(\ref{eq:energy-Schrodinger-2D}), versus $k_x$, with the $x$ axis chosen along $\hat{\bf k}$, so that $\xi = x$ and $\eta = y$.}
\label{Fig:energy-S}
\end{figure}

\begin{figure}
\includegraphics[width=0.9 \linewidth,angle=0] {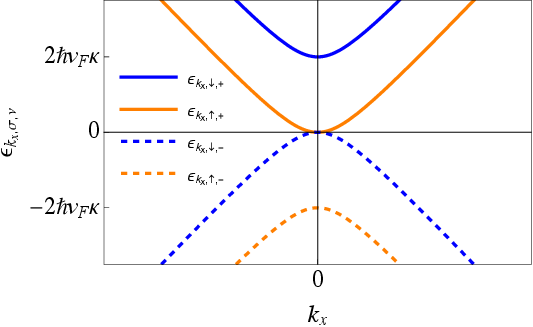}
\caption{\footnotesize
The eigenenergies $\epsilon_{k_x, \sigma, \nu}$ in Eq.~(\ref{eq:energy-Dirac}) of the Dirac Hamiltonian in Eq.~(\ref{eq:H_Dirac_2D}) versus $k_x$.  Here the $x$ axis is chosen along $\hat{\bf k}$, so $\xi = x$ and $\eta = y$.
$\kappa$ is given in the second equation of Eq.~(\ref{eq:kappa}).
}
\label{Fig:energy-D}
\end{figure}

It has already been demonstrated that Schr\"odinger equation (SE) for a particle with spin in the presence of an external electric field has an AC phase factor in 2D and 3D which is generally non-topological and non-Abelian \cite{Wu-Yang, Kuzmenko_25, Kuzmenko_25-long}.  Here we find that: (1) An analytical form exists for a plane wave solution to the SE for a particle with spin in the presence of an external electric field.  (2) The spin-dependent energy dispersion relation is parabolic as a function of band momentum, and its minimum shifts left and right from the $\Gamma$ point momentum (${\bf k} = {\bf 0}$) by an amount proportional to the strength of the external electric field for the two spin projection states (see Fig.~\ref{Fig:energy-S}). (3) The AC phase factor is Abelian, and the AC phase is geometrical.  (4) For single layer graphene modeled by the 2D Dirac equation (DE) \cite{Kane_05}, we show that the RSOI generates an AC phase for the electron wave function, and an effective electron mass term in the DE.  (5) The spin-dependent energy dispersion is hyperbolic as a function of the band momentum.  The minimum energy at the $\Gamma$ point is shifted up and down for the two spin projection states respectively (see Fig.~\ref{Fig:energy-D}).  (6) This splitting results in a time-dependent AC phase factor, hence the AC phase for the DE is dynamical.

{\it Aharonov-Bohm and Aharonov-Casher effects}---The Aharonov-Bohm (AB) effect \cite{AB_59} results in a phase of the wave function of an electrically charged particle (e.g., an electron) moving along a closed 1D curve (e.g., a planar circle) such that both the electric field ${\bf E}$ and the magnetic field ${\bf B}$ are zero along the curve.  The electron wave function is subject  to an electromagnetic $U(1)$ vector potential ${\bf A}$, such that the magnetic flux $\Phi$ through the area bounded by the curve is finite.
 
The AC effect \cite{AC_84} results in a phase of the wave function of a (not necessarily charged) particle with non-zero spin that moves along a closed 1D curve (e.g., a planar circle) such that the magnetic field along the curve vanishes, but the electric field ${\bf E} \ne 0$. This field generates a RSOI term in the Hamiltonian \cite{Rashba_59}.  For a spin 1/2 particle, it enters the Hamiltonian as an $SU(2)$ Rashba vector potential ${\bf A}_{R}=\alpha_R[{\hat{\bf E}} \times{\bm \sigma}]$, where ${\hat{\bf E}}$ is the unit vector in the direction of the external electric field, ${\bm \sigma}$ is the vector of Pauli matrices describing the spin $\uparrow, \downarrow$ \cite{AC_84, Bercioux_15}, and $\alpha_R$ is the RSOI coupling strength (in SI units, where $\alpha_R$ has units of are m/s),
\begin{equation}  \label{eq:Rashba_alpha}
  \alpha_R = \frac{g \mu_B E}{4 m_e c^2} ,
\end{equation}
where $g$ is the Land\'e $g$-factor, $\mu_B$ is the Bohr magneton, $m_e$ is the free electron mass, and $c$ is the speed of light in vacuum.  In general, $ \alpha_R$ depends on the material and on the confining potential \cite{Bercioux_15}.
 
It is well known that a $U(1)$ vector potential ${\bf A}$, as in the AB effect discussed in the previous paragraph, can be eliminated from a Hamiltonian  by a suitable gauge transformation, yielding the AB phase. A natural question is whether this elimination also applies for the  $SU(2)$ Rashba vector potential ${\bf A}_{R}$. We show below that it does not, indicating that determining the AC phase is more subtle.

{\it 2D Schr\"odinger equation with RSOI}---The Schr\"odinger Hamiltonian $H_{\rm S, 2D}$ of a 2D Fermi gas of electrons propagating in the $x$-$y$ plane in the presence of static homogeneous external electric field ${\bf E} = E \, \hat{\bf z}$ is \cite{Band-Avishai-Quantum-Mechanics}
\begin{equation}   \label{eq:H_S-2D}
  H_{\rm S, 2D} =
  \frac{{\bf p}^{2}}{2 m_e} \, \sigma_0 +
  \frac{\alpha_R}{2} \,
  \big\{
    {\bf p} \cdot [ {\boldsymbol \sigma} \times \hat{\bf z} ] +
    [ {\boldsymbol \sigma} \times \hat{\bf z} ] \cdot {\bf p}
  \big\} ,
\end{equation}
where ${\bf p} = p_x \, \hat{\bf x} + p_y \, \hat{\bf y}$ is the momentum operator, and $\sigma_0$ is the 2$\times$2 identity matrix acting in spin space.  The Hamiltonian in Eq.~(\ref{eq:H_S-2D}) can be rewritten as
\begin{equation}   \label{eq:H_S-2D-square}
  H_{\rm S, 2D} =
  \frac{1}{2 m_e} \, {\boldsymbol \Pi}^{2} -
  m_e \alpha_R^2 ,
\end{equation}
where the canonical momentum operator is
\begin{equation}   \label{eq:Pi}
  {\boldsymbol \Pi} =
  {\bf p} \, \sigma_0 + m_e \alpha_R \, {\boldsymbol \sigma} \times \hat{\bf z} .
\end{equation}
The constant term on the right hand side of Eq.~(\ref{eq:H_S-2D-square}), $-m_e \alpha_R^2$, simply shifts the energy.  It is important to note that the commutator of $\Pi_x$ and $\Pi_y$ is non-zero:
\begin{equation}
  \big[ \Pi_x, \Pi_y \big] = m_e^2 \alpha_R^2 \sigma_z .
\end{equation}
The RSOI cannot be eliminated from the Hamiltonian $H_{\rm S, 2D}$ in Eq.~(\ref{eq:H_S-2D-square}) by a unitary gauge transformation because no unitary operator can transform the non-commuting operators $\Pi_x$ and $\Pi_y$ into the commuting operators $p_x$ and $p_y$.  
Another way to show this is to observe that the energy eigenvalue $\epsilon_{k, \sigma}$ of $H_{\rm S, 2D}$ is
\begin{equation}  \label{eq:energy-Schrodinger-2D}
  \epsilon_{k, \sigma} =
  \frac{\hbar^2 k^2}{2 m_e} + \sigma \, \alpha_R \hbar k,
\end{equation}
where $\hbar {\bf k}$ is an eigenvalue of the electron momentum operator $\mathbf{p}$, $k = |{\bf k}|$, and $\sigma = \pm 1$ is the spin quantum number.  The energy $\epsilon_{k_x, \sigma}$ is plotted versus $k_x$ in Fig.~\ref{Fig:energy-S}, and is clearly parabolic in $k_x$, but its minimum is shifted in $k_x$ to the left and right of $k_x = 0$ for the states with $\sigma = \uparrow, \downarrow$ by $\pm \kappa$ (see  Fig.~\ref{Fig:energy-S} and Eq.~(\ref{eq:kappa})) respectively.  In the absence of an external electric field, $\alpha_R$ vanishes, therefore
\begin{equation}   \label{eq:energy^0}
\lim_{\alpha_R \to 0} \epsilon_{k, \sigma} = \epsilon_{k}^{(0)} = \frac{\hbar^2 k^2}{2 m_e} ,
\end{equation}
where $\epsilon_{k}^{(0)}$ is independent of $\sigma$.  No vector ${\boldsymbol \kappa}$ exists such that $\epsilon_{k, \sigma}$ can be written as $\epsilon_{{\bf k} + \sigma {\boldsymbol \kappa}}^{(0)}$, therefore the RSOI cannot be eliminated from the Hamiltonian $H_{\rm S, 2D}$ by a unitary transformation.

{\it Aharonov-Casher phase for a plane wave}---Although an AC phase factor that eliminates the RSOI cannot be defined for the 2D SE, the AC phase for a 2D plane wave function {\it can} be determined.  To this end, note that $H_{\rm S, 2D}$ commutes with the momentum operator ${\bf p}$, therefore the plane wave function
\begin{equation}   \label{eq:plane-wave-Schrodinger}
\psi_{{\bf k}, \sigma}({\bf r}) = \chi_{\hat{\bf k}, \sigma} \, e^{i {\bf k} \cdot {\bf r}}
\end{equation}
is an eigenfunction of $H_{\rm S, 2D}$, provided that the spinor $\chi_{\hat{\bf k}, \sigma}$ satisfies the equation,
\begin{equation}   \label{eq:for-chi-Schrodinger}
  \hat{\bf k} \cdot \big[ {\boldsymbol \sigma} \times \hat{\bf z} \big] \chi_{\hat{\bf k}, \sigma} =
  \sigma \, \chi_{\hat{\bf k}, \sigma} ,
\end{equation}
where the unit vector $\hat{\bf k}$ is
\begin{equation}   \label{eq:hat-k}
\hat{\bf k} =
\frac{\bf k}{k} =
\cos \theta_k \, \hat{\bf x} + \sin \theta_k \, \hat{\bf y} ,
\end{equation}
and $\theta_k$ is the angle that $\hat{\bf k}$ makes with the $x$ axis.  The operator on the left-hand side of Eq.~(\ref{eq:for-chi-Schrodinger}), $\hat{\bf k} \cdot [ {\boldsymbol \sigma} \times \hat{\bf z} ]$, can be written as ${\boldsymbol \sigma} \cdot \hat{\bf n}$, where $\hat{\bf n} \equiv \hat{\bf z} \times \hat{\bf k} = - \sin \theta_k \, \hat{\bf x} + \cos \theta_k \, \hat{\bf y}$.  The unit vectors $\hat{\bf k}$, $\hat{\bf n}$ and $\hat{\bf z}$ are perpendicular to one another.  Consequently, they can be used to define the basis of a Cartesian coordinate system $(\xi, \eta, z)$, where the $\xi$-axis is along $\hat{\mathbf{k}}$, and the $\eta$-axis is along $\hat{\mathbf{n}}$.
Solving the equation, ${\boldsymbol \sigma} \cdot \hat{\bf n} \, \chi_{\hat{\bf k}, \sigma} = \sigma \, \chi_{\hat{\bf k}, \sigma}$, for the eigenvector $\chi_{\hat{\bf k}, \sigma}$ of ${\boldsymbol \sigma} \cdot \hat{\bf n}$, and substituting the solution into Eq.~(\ref{eq:plane-wave-Schrodinger}), gives the 2D plane wave function,
\begin{equation}   \label{eq:psi_k_S}
  \psi_{{\bf k}, \sigma}({\bf r}) =
  \frac{1}{\sqrt{2}} \, \left( \begin{array}{c} 1 \\ i \, \sigma \, e^{i \theta_k} \end{array} \right) \,
  e^{i {\bf k} \cdot {\bf r}} ,
\end{equation}
and the eigenenergy $\epsilon_{k, \sigma}$ is given in Eq.~(\ref{eq:energy-Schrodinger-2D}).
Note that the wave function $\psi_{{\bf k}, \sigma}({\bf r})$ depends on $\xi = \hat{\bf k} \cdot {\bf r}$, but not on $\eta = \hat{\bf n} \cdot {\bf r}$. Therefore,  the electron motion is 1D, and the wave function $\psi_{{\bf k}, \sigma}({\bf r})$ is an eigenfunction of the following 1D Hamiltonian,
\begin{eqnarray}   \label{eq:H_S-1D}
  H_{\rm S, 1D} &=&
  \frac{1}{2 m_e} \, \Pi_{\xi}^{2} -
  \frac{m_e \alpha_R^2}{2}
  \nonumber \\ &=&
  \frac{1}{2 m_e} \, \big( p_\xi + m_e \alpha_R \, \sigma_n \big)^{2} -
  \frac{m_e \alpha_R^2}{2} ,
\end{eqnarray}
where $\Pi_\xi = p_\xi + m_e \alpha_R \, \sigma_n$, $\sigma_n = {\boldsymbol \sigma} \cdot \hat{\bf n}$, and $p_\xi = \frac{\hbar}{i} \frac{\partial}{\partial \xi}$.
There exists a unitary transformation with a unitary matrix ${\mathcal U}_{\hat{\bf k}}(\xi)$ such that the transformed 1D Hamiltonian takes the form,
\begin{eqnarray}   \label{eq:tilde-H_S-1D}
  \tilde{H}_{\rm S, 1D} &=&
  {\mathcal U}_{\hat{\bf k}}(\xi) \,
  H_{\rm S, 1D} \,
  {\mathcal U}_{\hat{\bf k}}^{\dag}(\xi) =
  \frac{p_\xi^2}{2 m_e} - m_e \alpha_R^2 .
\end{eqnarray}
The eigenenergies of $\tilde{H}_{\rm S, 1D}$ are spin-degenerate.  Therefore, the unitary matrix ${\mathcal U}_{\hat{\bf k}}(\xi)$ is an AC phase factor \cite{Kuzmenko_25, Kuzmenko_25-long, Wu-Yang}.  It can be rewritten as
\begin{equation}  \label{eq:AC-phase-factor-vs-AC-phase}
  {\mathcal U}_{\hat{\bf k}}(\xi) = e^{i \varphi_{\rm AC}^S(\xi) \, \sigma_n} ,
\end{equation}
where the AC phase $\varphi_{\rm AC}^S(\xi)$ is
\begin{equation}   \label{eq:AC-phase}
  \varphi_{\rm AC}^S(\xi) = \frac{m_e \alpha_R \xi}{\hbar} .
\end{equation}
Applying the unitary transformation to the wave function (\ref{eq:psi_k_S}) yields,
\begin{equation}\tilde\psi_{{\bf k}, \sigma}({\bf r}) = {\mathcal U}_{\bf k}(\xi) \, \psi_{{\bf k}, \sigma}({\bf r}) .
\end{equation}
Using Eq.~(\ref{eq:for-chi-Schrodinger}), the transformed wave function can be written as
\begin{equation}    \label{eq:plane-wave-Schrodinger_k_sigma}
  \tilde\psi_{{\bf k}, \sigma}({\bf r}) =
  \frac{1}{\sqrt{2}} \, \left( \begin{array}{c} 1 \\ i \sigma e^{i \theta_k} \end{array} \right) \,
  e^{i ({\bf k} + \sigma {\boldsymbol \kappa}) \cdot {\bf r}} =
  \psi_{{\bf k} + \sigma {\boldsymbol \kappa}, \sigma}({\bf r}) ,
\end{equation}
where
\begin{equation}    \label{eq:kappa}
  {\boldsymbol \kappa}= \kappa \, \hat{\bf k} , \quad \kappa = \frac{m_e \alpha_R}{\hbar} = \frac{g \mu_B E}{4 \hbar c^2}.
\end{equation}
$\tilde\psi_{{\bf k}, \sigma}({\bf r})$ is an eigenfunction of $\tilde{H}_{\rm S, 1D}$ with eigenenergy $\epsilon_{k, \sigma} = \epsilon_{k + \sigma \kappa}^{(0)} - \epsilon_{\kappa}^{(0)}$, see Eqs.~(\ref{eq:energy-Schrodinger-2D}) and (\ref{eq:energy^0}).  Note that the eigenenergy $\epsilon_{k, \sigma}$ depends on the electric field, through $\kappa$.

{\it Dirac equation with RSOI in graphene}---This section demonstrates that the RSOI in graphene gives rise to an AC phase of the electron wave function, and generates an effective electron mass.  Electrons in a single layer graphene lying in the $x$-$y$ plane, with momentum near the Dirac point satisfy a massless 2D DE.  Explicitly, the 2D Dirac Hamiltonian is \cite{Kane_05}
\begin{equation}   \label{eq:H_Dirac_2D}
  H_{\rm D, 2D} = v_F \, {\boldsymbol \tau} \cdot {\boldsymbol \Pi} ,
\end{equation}
where $v_F \approx 10^8$~cm/s is the Fermi velocity, ${\bm \tau}$ is a vector of Pauli matrices for the $A/B$ sub-lattices in single layer graphene, ${\boldsymbol \Pi} = \Pi_x \, \hat{\mathbf{x}} + \Pi_y \, \hat{\mathbf{y}}$ is the canonical momentum operator,
\begin{equation}   \label{eq:Pi-Dorac}
  {\boldsymbol \Pi} =
  {\bf p} \, \sigma_0 + \frac{\hbar e E}{4 m_e c^2} \, {\boldsymbol \sigma} \times \hat{\bf z} =
  {\bf p} \, \sigma_0 + m_e \, \alpha_R \, {\boldsymbol \sigma} \times \hat{\bf z} .
\end{equation}
As the canonical momentum operators $\Pi_x$ and $\Pi_y$ do not commute, whereas the momentum operators $p_x$ and $p_y$ do, the non-Abelian $SU(2)$ Rashba vector potential ${\bf A}_R = m_e \alpha_R \, {\boldsymbol \sigma} \times \hat{\bf z}$ cannot be eliminated by a unitary operator from the 2D Dirac Hamiltonian by a gauge transformation using a unitary operator.

The Hamiltonian $H_{\rm D, 2D}$ commutes with ${\bf p}$, therefore the plane wave function
\begin{equation}   \label{eq:plane-wave-Dirac}
  \psi_{{\bf k}, \sigma, \nu}(\xi) = \chi_{{\bf k}, \sigma, \nu} \, e^{i k \xi}
\end{equation}
is an eigenfunction of $H_{\rm D, 2D}$, provided that the 4-spinor $\chi_{{\bf k}, \sigma, \nu}$ satisfies the equation,
\begin{equation}   \label{eq:for-chi-Dirac}
  {\boldsymbol \tau} \cdot
  \Big\{
    \hat{\bf k} \, \sigma_0 +
    \frac{m_e \alpha_R}{\hbar k} \, {\boldsymbol \sigma} \times \hat{\bf z}
  \Big\}
  \chi_{{\bf k}, \sigma, \nu} =
  \frac{\epsilon_{k, \sigma, \nu}}{\hbar v_F k} \, \chi_{{\bf k}, \sigma, \nu} ,
\end{equation}
where  $\nu = \pm 1$ is the energy band number, $\xi = \hat{\bf k} \cdot {\bf r}$, and the unit vector $\hat{\bf k}$ is given in Eq.~(\ref{eq:hat-k}).
The eigenenergies of $H_{\rm D, 2D}$ are
\begin{equation}   \label{eq:energy-Dirac}
  \epsilon_{k, \sigma, \nu} =
  \hbar v_F \, \big( - \sigma \kappa + \nu \sqrt{k^2 + \kappa^2} \big) ,
\end{equation}
where $\kappa$ is given in Eq.~(\ref{eq:kappa}).
The Dirac eigenenergies $\epsilon_{k_x, \sigma, \nu}$ are plotted as a function of $k_x$ in Fig.~\ref{Fig:energy-D} for the case when the $x$ axis is taken to be along the unit vector $\hat{\bf k}$.  Note that the eigenenergies are shifted vertically here, not horizontally as in Fig.~\ref{Fig:energy-S} for the SE eigenenergies.

The wave function $\psi_{{\bf k}, \sigma, \nu}(\xi)$ in Eq.~(\ref{eq:plane-wave-Dirac}) is dependent on $\xi$, but not on $\eta = \hat{\bf n} \cdot {\bf r}$, where the units vectors $\hat{\bf z}$, $\hat{\bf k}$ and $\hat{\bf n} = \hat{\bf z} \times \hat{\bf k}$ are perpendicular to one another.  Therefore, $\psi_{{\bf k}, \sigma, \nu}(\xi)$ is an eigenfunction of the effective 1D Dirac Hamiltonian, $H_{\rm D, 1D}$, which can be derived from $H_{\rm D, 2D}$ in Eq.~(\ref{eq:H_Dirac_2D}) by substituting
${\boldsymbol \Pi} \to {\boldsymbol \Pi}_{\rm 1D} = ( p_\xi \sigma_0 + m_e \alpha_R \sigma_n) \, \hat{\bf k} - m_e \alpha_R \sigma_k \, \hat{\bf n}$:
\begin{equation}   \label{eq:H_Dirac_1D}
  H_{\rm D, 1D} =
  v_F \, \tau_k
  \big[ p_\xi \sigma_0 + m_e \alpha_R \, \sigma_n \big] -
  m_e v_F \, \alpha_R \tau_n \, \sigma_k .
\end{equation}
Here $\sigma_k = {\boldsymbol \sigma} \cdot \hat{\bf k}$, $\sigma_n ={\boldsymbol \sigma} \cdot \hat{\bf n}$, and similarly for $\tau_k$ and $\tau_n$.

{\it Time-dependent unitary transformation}---To compute the AC phase for the 1D DE, we must find a unitary transformation of $H_{\rm D, 1D} $ such that the transformed Hamiltonian has  spin-degenerate eigenenergies.  As will be shown below, this unitary transformation is {\it time-dependent}.  Therefore, we start with the time-dependent DE,
\begin{equation}   \label{eq:Dorac-time-dependent}
  i \hbar \, \frac{\partial \Psi(\xi, t)}{\partial t} = H_{\rm D, 1D} \Psi (\xi, t) .
\end{equation}
Let us apply a unitary transformation with a time-dependent unitary matrix
\begin{equation}
  {\mathcal U}_{\rm AC}(t) = \exp \big[ i \tau_k \sigma_n v_F \kappa \, t \big] .
\end{equation}
The Hamiltonian $H_{\rm D, 1D}$ in Eq.~(\ref{eq:H_Dirac_1D}) commutes with $\tau_k \sigma_n$, and therefore with ${\mathcal U}_{\rm AC}(t)$. The transformed (time-independent) Hamiltonian takes the form,
\begin{align} \label{eq:H_Dirac_1D_tilde}
  \tilde{H}_{\rm D, 1D} &\equiv
  {\mathcal U}_{\rm AC}(t) H_{\rm D, 1D} {\mathcal U}^\dagger_{\rm AC}(t) -
  i \, \hbar \, {\mathcal U}_{\rm AC}(t) \, \frac{\partial {\mathcal U}_{\rm AC}^{\dag}(t)}{\partial t}
  \nonumber \\ &=
  v_F p_\xi \, \tau_k \sigma_0 -
  \hbar v_F \kappa \, \tau_n \sigma_k .
\end{align}
The eigenenergy $\tilde\epsilon_{k, \nu}$ of $\tilde{H}_{\rm D, 1D}$ is
\begin{equation}   \label{eq:eepsilon^0-D}
  \tilde\epsilon_{k, \nu} = \hbar v_F \nu \, \sqrt{k^2 + \kappa^2} .
\end{equation}
$\tilde\epsilon_{k, \nu}$ are spin-degenerate.  The unitary transformation does {\it not} eliminate the RSOI from the Dirac Hamiltonian, since the second term on the right-hand side of Eq.~(\ref{eq:H_Dirac_1D_tilde}) depends on the electric field through $\kappa$, and generates an effective electron mass $\hbar v_F \kappa$.  Since the spin-splitting is eliminated, the unitary matrix ${\mathcal U}_{\rm AC}(t)$ is the AC phase factor.  The AC phase factor can be rewritten as
\begin{equation}   \label{eq:U_AC^D}
  {\mathcal U}_{\rm AC}(t) =
  \exp \big( i \, \tau_k \sigma_n \, \varphi_{\rm AC}^D(t) \big) ,
\end{equation}
where
\begin{equation}   \label{eq:phi_AC^D}
  \varphi_{\rm AC}^D(t) = v_F \kappa \, t .
\end{equation}
The 4$\times$4 matrix $\tau_k \sigma_n$ (i.e., $\tau_k \otimes \sigma_n$) has two doubly-degenerate eigenvalues, $S_1 = S_2 = 1$, and $S_3 = S_4 = -1$.  Therefore, ${\mathcal U}_{\rm AC}(t)$ has two eigenvalues, $e^{\pm i \varphi_{\rm AC}^D(t)}$. The AC phase, $\varphi_{\rm AC}^D(t)$, is {\it time-dependent}; making it a dynamical phase.

{\it Summary and Conclusions}---We have shown that the RSOI term cannot be eliminated from either the 2D Schr\"odinger equation or the 2D Dirac equations via a gauge transformation.  However, an $SU(2)$ matrix ${\mathcal U}_{\hat{\bf k}}(\xi)$ exists that eliminates the RSOI term $m_e \alpha_R \sigma_n$ from the 1D Schr\"odinger Hamiltonian in Eq.~(\ref{eq:H_S-1D}). The transformed Hamiltonian, $\tilde{H}_{\rm S, 1D}$ in Eq.~(\ref{eq:tilde-H_S-1D}), has spin-degenerate eigenenergies.  The AC phase factor, ${\mathcal U}_{\hat{\bf k}}(\xi)$, as well as the AC phase, $\varphi_{\rm AC}^S(\xi)$, depend on the coordinate $\xi$; therefore $\varphi_{\rm AC}^S(\xi)$ is a geometrical phase.  For a plane wave, the Dirac 1D Hamiltonian $H_{\rm D, 1D}$ in Eq.~(\ref{eq:H_Dirac_1D}) contains two terms proportional to the electric field strength: $m_e v_F \alpha_R \tau_k \sigma_n$ and $-m_e v_F \alpha_R \tau_n \sigma_k$.  The latter generates an effective electron mass $m_e v_F \alpha_R = \hbar v_F \kappa$, that cannot be eliminated from the Dirac Hamiltonian via a unitary transformation.  The former generates an AC phase that can be eliminated by a time-dependent unitary transformation ${\mathcal U}_{\rm AC}(t)$ [see Eq.~(\ref{eq:U_AC^D})]. Thus, the Dirac AC phase $\varphi_{\rm AC}^D(t)$ in Eq.~(\ref{eq:phi_AC^D}) is {\it time-dependent}, that is, it is a dynamical phase.  This is in contrast to the Schr\"odinger AC phase which is geometrical.  Hence, the nature and characteristics of the AC phase is more subtle than previously realized.


\end{document}